# Photophoretic Levitation of Macroscopic Nanocardboard Plates

**Authors:** John Cortes[1], Christopher Stanczak[2], Maanav Narula[1], Mohsen Azadi[1], Samuel M. Nicaise[1], Howard Hu[1], and Igor Bargatin[1]*

**Affiliations:** [1]Department of Mechanical Engineering and Applied Mechanics, University of Pennsylvania, Philadelphia, PA. [2]Vagelos Integrated Program in Energy Research, University of Pennsylvania, Philadelphia, PA.

*Correspondence and requests for materials should be addressed to bargatin@seas.upenn.edu.

**Summary:** Scaling down miniature rotorcraft and flapping-wing flyers to sub-centimeter dimensions is challenging due to complex electronics requirements, manufacturing limitations, and the increase in viscous damping at low Reynolds numbers. Photophoresis, or light-driven fluid flow, was previously used to levitate solid particles without any moving parts but only at the scale of 1-20 micrometers. Here, we leverage stiff architectured plates with 50-nm thickness to realize photophoretic levitation at the millimeter to centimeter scales. Instead of creating lift through conventional rotors or wings, our levitation occurs due to light-induced thermal transpiration through micro-channels within our plates. At atmospheric pressure, the plates hover above a solid substrate at heights of ~0.5 mm by creating an air cushion beneath the plate. Moreover, at reduced pressures (10-1000 Pa), the increased speed of thermal transpiration through the plate's channels creates an air jet that enables mid-air levitation and allows the plates to carry small payloads heavier than the plates themselves. This new propulsion mechanism for macroscopic structures creates the prospect of future nanotechnology-enabled flying vehicles without any moving parts in the Earth's upper atmosphere and at the surface of other planets.

Miniaturized unmanned aerial vehicles or drones typically rely on conventional rotorcraft or flapping wing propulsion, with recent advances made possible by new actuators, processors, batteries, sensors, and communication electronics[1-3]. The current record holder for the smallest self-powered maneuverable drone is the Piccolissimo, which is about 28 mm in its largest dimension, weighs approximately 2.5 grams, and uses rotorcraft propulsion[4]. Another approach is provided by winged robots inspired by insects The RoboBee, which weighs only ~80 mg and a few cm in size, has demonstrated the ability to lift off, hover and move in the horizontal plane using power tethers or concentrated sunlight[5,6].

Many emerging applications for aerial vehicles, such as search and rescue missions, jet engine inspection, and even satellites could benefit from even smaller, "smart dust"-scale flyers[7]; however, further decreasing the size of conventional microflyers introduces many challenges. One key challenge is the increasing difficulty of miniaturizing both mechanical structures and electronic components. Another, more fundamental challenge is the fact that the propulsion efficiency of rotorcraft and winged flyers decreases as the vehicle scale is reduced and Reynolds number decreases, making viscous forces more dominant[8]. These limitations can be overcome by using alternative methods of propulsion, such as the photophoretic forces described below.

Solids illuminated by light while immersed in a gas experience a force variously known in different literature domains as the photophoretic, radiometric, or Knudsen force[9]. Different from the light pressure caused by a photon's momentum, this force appears when incident light is absorbed by a solid, heating the solid relative to the ambient gas. The temperature difference can then drive gas flows that result in a corresponding reaction force acting on the solid. The photophoretic/radiometric/Knudsen force has been studied for over a century, including by luminaries such as Reynolds, Maxwell, Einstein, and Knudsen, who primarily focused on explaining a common device called a light mill or Crookes radiometer[10,11]. The force pushing on the rotating vanes of a light mill is typically a few orders of magnitude smaller than the weight of the centimeter-scale paper vanes, which is enough to produce rotation on a low-friction bearing, but not for levitation. However, for micrometer-sized particles, the same force can exceed their weight. Both experimental[12,13] and theoretical studies[14] have shown the levitation of 1-to-20 micrometer-sized spherical particles in mid-air by using light to create a temperature difference and thus gas flow across the particle surface. The corresponding movements of particles and gas are called photophoresis, from Greek words for light and motion.

Much of the recent research on photophoretic forces focused on the settling of aerosols in the atmosphere[15,16] and led to a better understanding of how pollutants such as volcanic ash, carbon soot and other greenhouse gases affect atmospheric conditions[17]. Recently, researchers have also modeled the photophoretic lift forces generated on aerosol nano-particles as a function of altitude, which could be useful for climate engineering efforts[18,19]. Finally, some of the most detailed theoretical models of photophoretic forces have been developed to study the effect of photophoresis on planetary disk accretion[20]. On the more applied side, Smalley et al. recently created a 3D display by photophoretically trapping and moving 10-micrometer cellulose particles in free space while continuously illuminating them with red, green and blue light, providing highly detailed three-dimensional images[21]. Yet, despite multiple proposals in the literature[22-24], sustained levitation of larger, macro-scale bodies has not yet been demonstrated because it requires materials with both very low density and thermal conductivity. This creates an opportunity for the development of new, engineered materials coupled with photophoresis to realize new flying devices.

While the previous research on photophoresis focused on the gas flows around solid bodies, such as spheres or plates, photophoretic flows can also occur inside tubes and channels if a temperature gradient is present along the inner channel wall, as shown in Fig. 1B. This phenomenon, known as thermal transpiration or thermal creep[28], has been widely studied both fundamentally[25-27] and with the goal of creating new types of gas pumps without any moving parts[28,29]. As we show below, thermal transpiration can also be used to pump air through an architected plate to maximize the photophoretic forces and realize photophoretic hovering and propulsion of macroscopic structures at both atmospheric and reduced pressures.

The physical mechanism behind the photophoretic force depends on the Knudsen number, $Kn = \lambda/L$, where $\lambda$ is the molecular mean free path, and $L$ the characteristic length scale of the flow. For reference, the mean free path for air at atmospheric pressure and room temperature is ~70 nm and scales proportionally to the absolute temperature and inversely proportionally to the pressure[30]. In the free-molecular regime ($Kn \gg 1$), the gas molecules that collide with the hotter surface depart at a larger thermal velocity in comparison to the molecules colliding with the cooler surface. This momentum exchange results in a photophoretic force that is proportional to the number of molecules colliding with the solid per unit time and, therefore, proportional to both the active area and the pressure of the ambient gas. In contrast, in the continuum regime ($Kn \ll 1$), the ambient gas will flow along the side walls of the object due to the thermal creep phenomenon[18], producing a reaction force that usually scales with the perimeter length of the plate and inversely proportionally to the pressure (see Supplementary Information for details). As a result, the force is typically maximum at pressures that correspond to the transition regime ($Kn \sim 1$). In all these regimes, the resulting force is directed from the hot side of the device towards the cold side, meaning that the incident light should be primarily absorbed by the bottom side of the solid to create photophoretic lift.

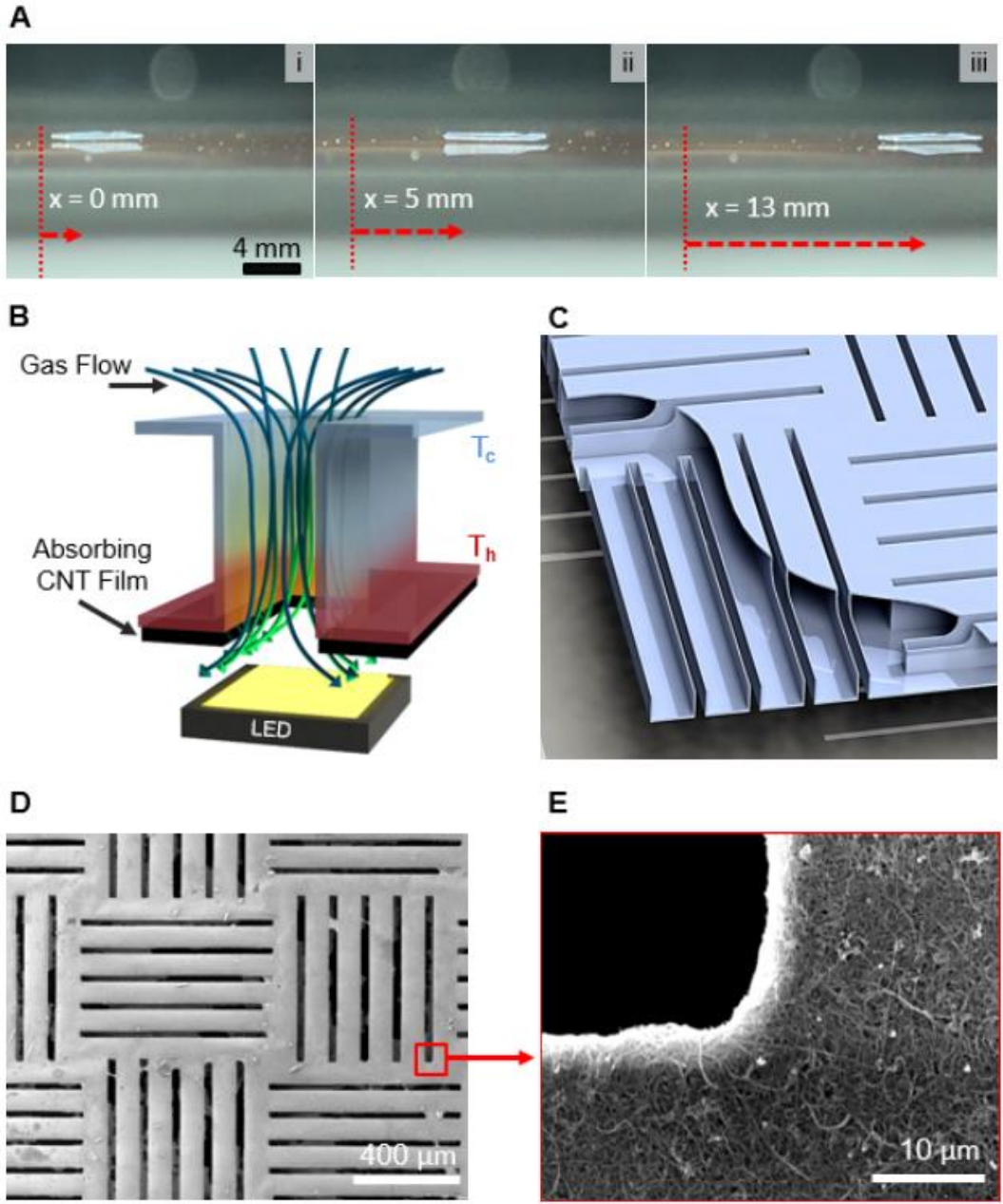

Figure 1. (A) Sequential screenshots of a low-height levitation test at atmospheric conditions inside of a glass enclosure at an incident optical flux of 10,000 W/m$^2$ (~10 Suns). (B) Schematic of the photophoretic force-induced gas flow through one nanocardboard channel. (C) Cross sectional schematic of the optimized nanocardboard structure, showing a cutaway view of the internal channels. (D) Scanning electron micrograph (SEM) of the 5-channel-array optimized nanocardboard plate structure. (E) SEM micrograph of the 200 nm thick carbon nanotube film on the edge of one of the nanocardboard channels.

Fig. 1 shows an example of our previously reported nanocardboard[31] — a microarchitected metamaterial that is designed to be light enough photophoretic flows through the engineered channels can levitate the plates without the need of moving parts. This particular design features arrays of 5 channels, with in-plane dimensions of 25 × 600 µm and 75-µm gaps (Fig. 1D), and results in air flowing through the plate at an effective "flow-though" velocity on the order of 1-10 mm/s. Though these cm-scale plates are ultralight by nature (~1 g/m$^2$), they are robust enough to handle with tweezers and even by hand without causing permanent damage.

These ultralight plates were fabricated on a 4-inch silicon on insulator (SOI) wafer, with silicon device layers varying between 10 µm and 200 µm. Photolithography was used to pattern the device layer, followed by deep reactive-ion etching (DRIE) to create the channel openings. After the device layer was separated from the carrier wafer, an alumina was conformally coated via atomic layer deposition (ALD). This alumina layer, typically 50-nm thick, coated all exposed surfaces, including the channel walls. A thin carbon nanotube (CNT) film was then deposited by drop-casting on one side of the plate and selectively etched in oxygen plasma to form the absorbing layer (as seen in Fig. 1D and 1E). The samples were then cleaved to expose the encapsulated silicon, then etched chemically in xenon difluoride (XeF$_2$) gas. The resulting hollow plates of alumina were manually cut with a razor blade to the desired size (typically, ~1 cm). For more details of the fabrication process, see the supplementary information's main text.

Fig. 1A and Supplementary Movie 1 show our plates hovering above a transparent substrate at equilibrium heights ranging from 400 to 600 µm for arbitrarily long periods of time. The levitation was most reliably observed above a micropatterned low-stiction glass substrate (see Supplementary Information main text). We observed the hovering in over 50 flights with a wide variety of samples. The levitation heights were estimated from microscope video footage, such as the one shown in Fig. 1A, by comparing the distance between the plate and its reflection to the 500-µm thickness of the substrate. The incident optical flux of up to ~1 W/cm$^2$ was produced by a 100-Watt LED diode array, as shown in Fig. 2B. The light was absorbed by the CNT film on the underside of the plate, causing the temperature gradient along the channels and therefore the pumping of air through the channels into the gap below the plate, creating an air cushion just like in a conventional hovercraft.

To predict the lift force as a function of the levitation height, we developed a theoretical model based on the well-known lubrication theory for thin-film flows under a disk with an imposed velocity through it (see Supplementary Information main text). As discussed in the Supplementary Information, the lift force on a levitating plate is typically proportional to the total volumetric flow rate of the gas through plate's channels. Therefore, the structure shown in Fig. 1C has the number and size of channels chosen to maximize the ratio of the gas flow through the plate to its weight.

Fig. 2D shows the theoretical equilibrium height, at which the lift force is equal to the weight of the plate, for both single-channel and optimized 5-channel nanocardboard plates. The theoretical levitation height for these 5-channel optimized plates is ~500 µm, which agrees with the experimentally observed heights. One potential reason for the discrepancy between the tests and theoretical model is that our model assumes perfectly flat plates whereas, in reality, the plates curve upwards near the edges due to differential thermal expansion. Another potential source of error is the inconsistency in thickness of the absorbing carbon nanotube layer, which can vary by as much as 50 nm for a nominal thickness of 200 nm. This inconsistency results in both weight and absorption variations that ultimately affect the plate's ability to generate the gas flow through the plate.

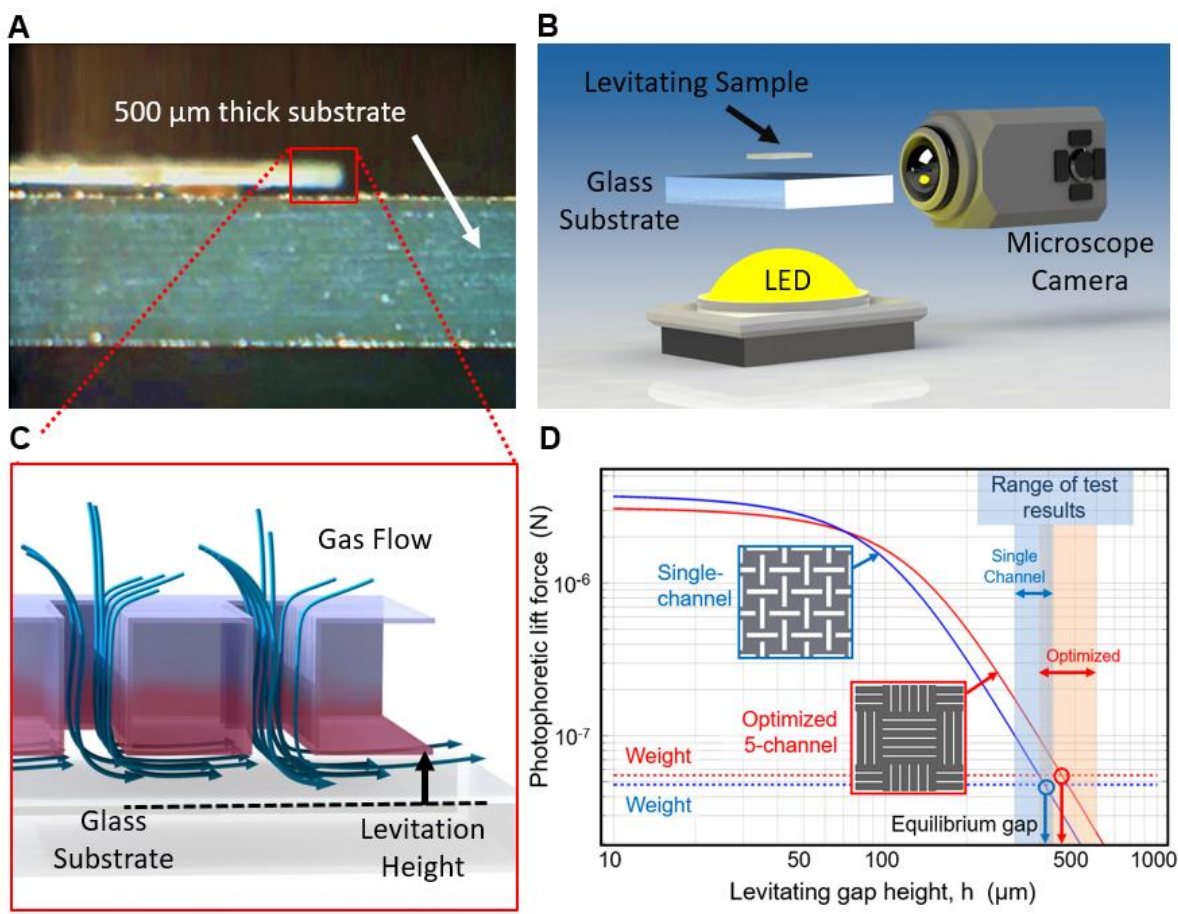

Figure 2. (A) Microscope camera picture showing a 60 µm-thick sample levitating at a small height above a 500 µm-thick substrate, which is used as a measuring reference (B) Graphical schematic of the image acquisition setup (not to scale). (C) Schematic of the levitation mechanism at atmospheric pressure based on thermal creep flows (not to scale) (D) Prediction of the equilibrium gap height for nanocardboard plates with single and 5-channel designs levitating above a substrate based on our theoretical model for the incident optical flux of 1 W/cm$^2$ (~10 Suns).

For highly insulating materials in the continuum regime, the thermal creep velocity and the flow-through velocities are on the order of $I/(10P)$, where $I$ is the incident light flux and $P$ is the ambient pressure[25,32]. Therefore, the volumetric flow rate through our plates can be increased dramatically by reducing the ambient pressure, resulting not only in hovercraft-like levitation on an air cushion but also mid-air flight. Fig. 3A and Supplementary Movie 2 show sequential screenshots of a plate in mid-air flight at a reduced pressure of approximately 1 kPa inside a glass vacuum chamber. Unlike the hovering behavior previously discussed where the gaps underneath the plate were less than 1 mm, these flights featured heights as high as 10 mm even though we replaced the solid glass substrate with a very sparse metal mesh (84% open area) to eliminate the air cushion effect. Instead of the air cushion, the lift force was generated by the jet of air going through the plate at a flow-through velocity of ~1 m/s (Fig 3B). As described in the Supplementary Information, the lift force can be predicted by calculating the drag force for thin plates in a certain moving frame of reference. Alternatively, the ability of a plate to levitate can be predicted by comparing the terminal velocity of a freefalling plate to the flow-through velocity, i.e., the average velocity of the air pumped through the plate by thermal transpiration.

Fig. 3C below shows a comparison of the photophoretic lift force and a plate's weight as a function of pressure for an optimized plate measuring 0.5 cm × 1 cm. The lift force exceeds the weight in a wide pressure range from 0.2 to 200 Pa, which agrees with our experimental results for plates of this size and geometry. We observed no levitation at pressures above 200 Pa, while pressures between 20 and 200 Pa resulted in reliable levitation (the glass vacuum chamber we used did not allow vacuum pressures below ~20 Pa). The total time for the levitation was typically less than 1 second as the plates would typically move horizontally and escape from the light beam created by the LED array.

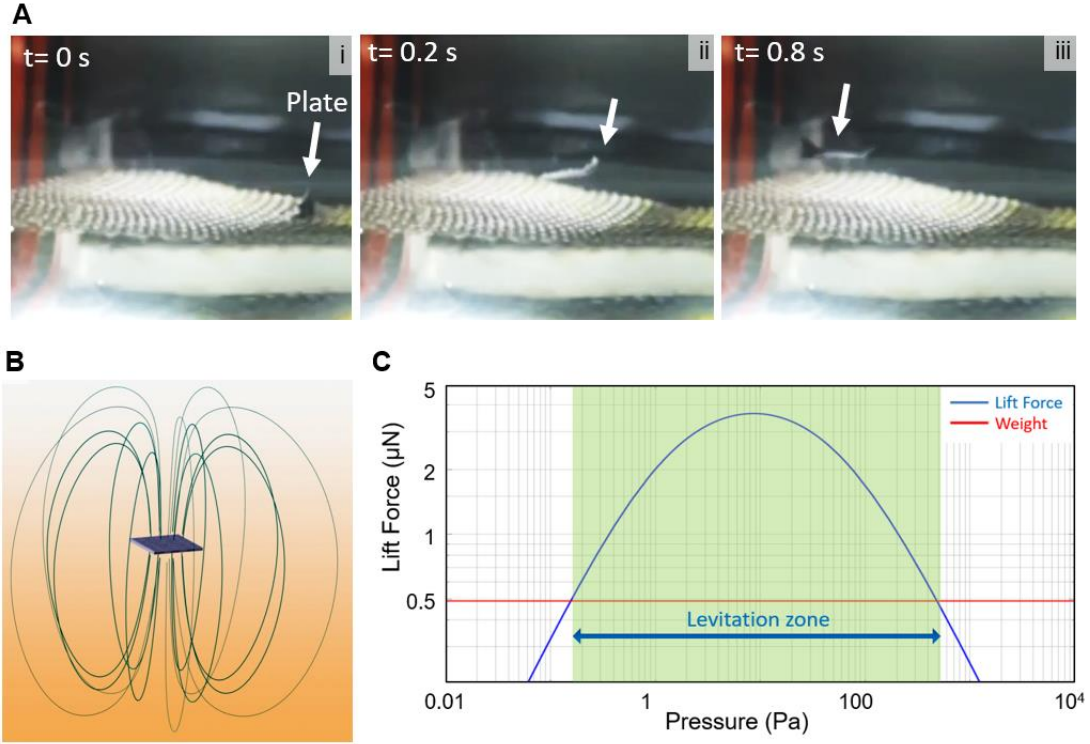

Figure 3. A(i→ ii→ iii) Sequential screenshots of a levitation test at reduced pressures inside of a vacuum chamber at an illumination of 0.8 W/cm$^2$ (~8 Suns). The plate being tested has dimensions of approximately 0.5 cm × 1 cm. (B) Schematic of the flow fields around a plate levitating in mid-air, based on numerical flow simulations. (C) Theoretical model prediction of the pressure regimes under which mid-air levitation is possible, based on a comparison of the lift force and weight for a plate measuring 0.5 cm × 1 cm at an illumination of 1 W/cm$^2$ (~10 Suns).

Finally, we explored the payload capability of the nanocardboard plates at reduced pressure conditions by lifting and carrying small silicon rings. The typical rings had an inner radius of 950 µm, an outer radius of 1000 µm, a height of 200 µm, corresponding to a mass of ~0.3 mg. Fig. 4A shows the plate with the silicon ring on top of it as well as a set of sequential screenshots from a test flight at a pressure of ~100 Pa. In Fig. 4A-ii, the nanocardboard plate lifted off from the metal mesh, then drifting towards the right side of the frame (Fig. 4A-iii). In this test, the mass of the of the plate alone was approximately 100 µg, while the payload weighed approximately 330 µg.

Fig. 4B shows theoretical predictions for the maximum areal density of the plates that can be levitated as a function of their side length and the ambient pressure. At the standard sea-level atmospheric pressure (~100 kPa), even sub-millimeter plates need to have an areal density of less than 0.1 g/m$^2$, which is below the lightest plates we manufactured (~1 g/m$^2$), explaining why we did not observe mid-air levitation at the atmospheric pressure. However, at reduced pressures, the lift force can exceed the weight of the plate by many orders of magnitude. Moreover, photophoretic propulsion always works better as the devices get smaller, as higher areal densities can be levitated as plate size is decreased (Fig. 4B), in contrast to conventional propulsion methods.

Fig. 4C shows the theoretical predictions for the payloads that can be levitated by our plates. For example, a 3 cm × 6 cm plate has the theoretical ability to lift as much as 2 mg at a pressure of approximately 10 Pascals and an optical flux of 1 W/cm$^2$ (10 Suns). The model's predictions agree with the experimental we performed, such as the one shown in Fig. 4A. Much progress has recently been made in creating such milligram-scale payloads for "smart dust" technology[33], with multiple demonstrations of electronic systems with cubic-millimeter dimensions and few-mg mass, such as fully functional sensing platforms[34,35] and aerospace applications[36]. Recent research on ultrathin electronics and solar cells[37-39] provides another avenue to self-powered ultralight payloads with multiple functionalities.

The levitation and payload capabilities of our plates peak in the 1-100 Pa range, which is ideal for applications in the upper atmosphere of Earth or on Mars. For example, in the Earth's mesosphere, which corresponds to altitudes from 50 to 80 km, the air is too thin for conventional airplanes or balloons but too thick for satellites, so aviation-based measurements can be performed for only a few minutes at a time during the short flight of a research rocket. However, the range of ambient pressures in the mesosphere (1-100 Pa) is nearly optimal for our plates' payload capabilities. With further optimization, the photophoretically levitated micro-flyers could stay aloft for extended periods of time by using the Sun's natural light throughout the daytime, or even indefinitely if the micro-flyer system is designed to ascend during the day and descend the same vertical distance at night. Microflyers based on photophoretic propulsion would enable new ways to study the mesosphere using micro-sensor payloads that weigh only a few milligrams. Similarly, the Martian atmosphere has a pressure of about 600 Pa at the surface, making photophoretic microflyers an excellent fit for exploring Mars and other celestial bodies with similarly thin atmospheres. The microflyers could work alongside current terrestrial rovers to study the Martian atmosphere and provide a cheaper, small-scale alternative to the Mars helicopter[40].

While this paper demonstrated the proof of concept of photophoretic propulsion for macroscopic structures, the lift force can be greatly increased in the future by further optimizing the geometry and using advanced optical absorbers. By placing optical absorbers with a large optical cross section, such as nanoparticles, at the entrances of the channels, local temperature gradients can be enhanced by at least an order of magnitude. Because the speed of the thermal-transpiration jet is proportional to a local temperature gradient, resonant optical absorbers film can lead to much more efficient levitation than the carbon nanotube films we used as uniform, broadband absorbers here. The use of localized absorbers would also allow for a new optimized design which features even more channels per unit area, since the absorption is no longer as dependent on the geometric fill factor. The corresponding increase in the lift force would enable much larger payloads, as well as levitation at the regular sea-level pressure (~100 kPa). Optical metamaterials can also be used to create local resonant absorption and facilitate the control of the flight[41]. Future research will focus on such improvements in the lift force as well as effective control of the flight.

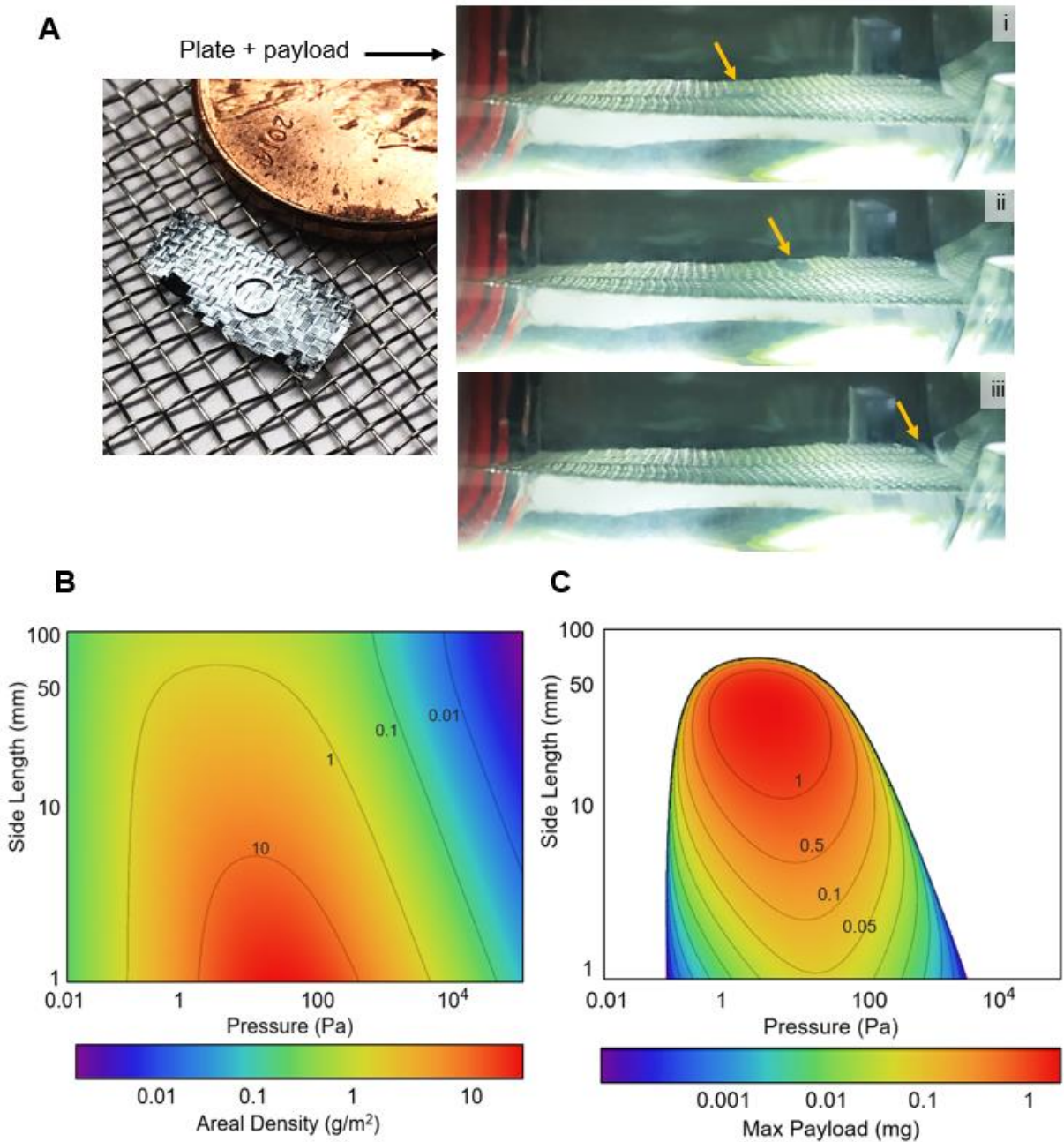

Figure 4. A(i→ ii→ iii) Sequential screenshots of a levitation test consisting of a nanocardboard rectangular plate (6×13 mm) carrying a microfabricated silicon payload (shown sitting on top of the plate) at an illumination of about ~10 Suns. (B) Density plot of the areal density that can levitate as a function of pressure and plate width (assuming rectangular plates with 2:1 aspect ratio). (C) Density plot of the maximum payload as a function of pressure and plate side length (assuming nanocardboard plate with areal density of 1 g/m$^2$).

**Methods:**

Plate fabrication process

The ultralight plates were fabricated on a 4-inch silicon on insulator (or SOI) wafer, with silicon device layers varying between 10 µm and 200 µm. A 3 µm layer of silicon oxide was deposited through plasma-enhanced chemical vapor deposition (PECVD) on top of the device layer to serve as the hard mask during the deep etching process. Microposit S1818 photoresist was spun to a height of 2.6 µm. then soft-baked at 115º C for 1 minute. Next, the channel-array pattern was directly written onto the photoresist using a Heidelberg 66+ laser writer at 250*mW* with a multi-pass process to ensure proper exposure and pattern continuity.

After photoresist development, the hard mask was etched in a Reactive Ion Etching (RIE) process of Trifluoromethane and Oxygen ($CHF_3/O_2$) chemistry and power is maintained at 150W. The silicon device layer was then etched in a deep reactive ion etcher (DRIE) to create the channel openings, as seen in Fig. S1A. Next, the device layer was separated from the carrier wafer through HF etching of the buried oxide layer so that only the perforated silicon mold remained, as shown in Fig S1B. A thin, conformal layer of aluminum oxide, or alumina was deposited through atomic layer deposition (ALD) at 250 C, using trimethylaluminum and water as the process precursors. The thickness, ranging from 25 nm to 100 nm, was controlled by changing the number of sequential deposition cycles performed. The thin layer is illustrated, in blue, in Fig. S1C.

After the alumina was deposited, a dilute solution of carbon-nanotubes (CNTs) in DI water was sprayed on the top surface of the disks. Once the water evaporated, the carbon nanotubes remained on the surface, leaving a film that covered a large portion of the channel openings (Fig. S1E). In order to open the channels so that gas could flow through the nanocardboard plate, the carbon nanotube film needed to be selectively removed. This was done by placing the alumina-covered silicon disks upside down in a reactive ion etcher and by using oxygen plasma to remove the carbon nanotubes that clog the opening, resulting in completely open channels, as shown in Fig. S1F. The samples were then cleaved, exposing the trapped silicon, which was then etched chemically through xenon difluoride ($XeF_2$) gas, leaving only the aluminum oxide shell shown in Fig. S1D. The resulting plates were manually cut to the desired size for levitation testing.

Testing methods

The levitation tests at atmospheric pressure were performed in an enclosed glass chamber measuring approximately 10"×6"×7", as shown in Fig. S2A. The enclosed chamber is important because strong air conditioning currents can pick up and carry our ultralight samples. The images and videos were captured by a microscope camera that featured an adjustable focal plane. The optical flux required for the lift force generation was provided by a 100-Watt LED diode which can produce up to ~4 W/cm$^2$ = 40 000 W/m$^2$, or 40x the Sun's flux on earth (about 1 000 W/m$^2$ on a clear sunny day). The LED diode was connected to a heatsink through a set of heat pipes to ensure safe and efficient operating temperatures.

The plates were initially placed on an engineered low-stiction substrate (as shown in Fig. S2B), which consisted of a fused-silica wafer featuring a 200 nm thick layer of gold that was patterned to minimize contact stiction. The honeycomb gold pattern (see Fig. S2B inset) had ribs with a width of approximately 2 µm. The low contact surface between the substrate and the plates ensured that attractive van der Waals molecular forces were minimized. Though contact stiction was the primary obstacle to observing lift-off, we found that electrostatic forces also played a role. To combat these effects, we grounded the gold film minimizing electrostatic forces. The sparse gold pattern ensured that minimal optical flux was reflected or absorbed by the gold layer, with the measured reduction in flux of only 6-7%.

Testing at reduced pressures required the use of a transparent vacuum chamber, which had a diameter of approximately 4" and a depth of 5", as shown in Fig. S2C. The chamber was connected to a roughing pump able to evacuate the system to pressures as low as ~10 Pa. Testing pressures were monitored by a gauge situated parallel to the vacuum line. Once the desired pressure was reached, a control valve was closed to ensure that the chamber remained at that pressure during flight testing. An aluminum grid with an 84% open area (shown in Fig. S2D) was used to lessen both the stiction and the ground effect.

**Acknowledgments:** The authors thank the staff of the Singh Center for Nanotechnology, Nanoscale Characterization Facility and Scanning and Local Probe facility at the University of Pennsylvania, which were partly funded by the NSF National Nanotechnology Coordinated Infrastructure Program, under grant NNCI-1542153. Special thanks to Dr. Hiromichi Yamamoto, Dr. Gerald Lopez, Meredith Metzler and David Jones. This work is supported in part by the United States Department of Education Graduate Assistance in Areas of National Need (GAANN) Fellowship (J.C.), the National Science Foundation under CBET-1845933, and the School of Engineering and Applied Science at the University of Pennsylvania.

**Author contributions:** The experimental samples were fabricated by J.C. Experimental testing and characterization was performed by J.C and C.S. and M.A. Theoretical models were developed by J.C., I.B., and H.H. Numerical simulations were performed by C.S. and M.N. The project was conceived and directed by I.B. The manuscript was written by J.C. and I.B. with contributions from all authors.

**Competing interests:** The authors declare no competing interests.

**Data and materials availability:** All relevant data and code are available from the authors upon request.

**Supplementary Information:**

Supplementary Text

Figures S1-S13

Movies S1-S2